%% file: alicepreprint_CDS.tex
\documentclass[ALICE,manyauthors]{cernphprep}
\usepackage{hyperref}
\usepackage[comma,square,numbers,sort&compress]{natbib}
\usepackage{amsmath}
\usepackage{graphicx}
\usepackage{dcolumn}
\usepackage{bm}
\usepackage{xspace}
\usepackage[para]{threeparttable}

\begin{document}%

\newcommand{\pp}{\ensuremath{\rm pp}\xspace}
\newcommand{\ks}{\ensuremath{\rm K_{S}^{0}}\xspace}
\newcommand{\lmd}{\ensuremath{\Lambda}\xspace}
\newcommand{\cme}{\ensuremath{\sqrt{s}}\xspace}
\newcommand{\bg}{\ensuremath{\beta\gamma}\xspace}
\newcommand{\pt}{\ensuremath{p_{\rm{T}}}\xspace}
\newcommand{\p}{\ensuremath{p}\xspace}
\newcommand{\gev}{\ensuremath{{\rm GeV}/c}\xspace}
\newcommand{\gevc}[1]{\ensuremath{#1 \text{ GeV/$c$}}\xspace}
\newcommand{\pbpb}{Pb--Pb\xspace}
\newcommand{\auau}{Au--Au\xspace}
\newcommand{\ptopi}{\ensuremath{({\rm p+ \bar{p}}) / (\pi^{+}+\pi^{-})}\xspace}
\newcommand{\ktopi}{\ensuremath{({\rm K}^{+}+{\rm K}^{-}) / (\pi^{+}+\pi^{-})}\xspace}
\newcommand{\twotwo}{\ensuremath{2\rightarrow 2}\xspace}
\newcommand{\raa}{\ensuremath{R_{\rm AA}}\xspace}
\newcommand{\dedx}{\ensuremath{{\rm d}E/{\rm d}x}\xspace}
\newcommand{\mdedx}{\ensuremath{\langle {\rm d}E/{\rm d}x \rangle}\xspace}
\newcommand{\pikp}{\ensuremath{\pi^{\pm}, \rm K^{\pm}, \rm p(\bar{p})}\xspace}

%
%
\begin{titlepage}
\PHyear{2013}
\PHnumber{230}      
\PHdate{15 Dec}
%
%
\title{Production of charged pions, kaons and protons at large transverse momenta in {\bf pp} and \pbpb collisions at $\mathbf{\sqrt{s_{\rm \textbf{NN}}}=2.76}$ TeV}
\ShortTitle{Identified particle production at high \pt in \pbpb collisions}   
%
\Collaboration{ALICE Collaboration%
         \thanks{See Appendix~\ref{app:collab} for the list of collaboration
                      members}}
\ShortAuthor{ALICE Collaboration}      
\begin{abstract}
Transverse momentum spectra of $\pi^{\pm}$, $\rm K^{\pm}$ and $\rm
p$($\bar{\rm p}$) up to $\pt = \gevc{20}$ at mid-rapidity in \pp, peripheral
(60-80\%) and central (0-5\%) \pbpb collisions at $\sqrt{s_{\rm NN}} =$
2.76\,TeV have been measured using the ALICE detector at the Large Hadron
Collider. The proton-to-pion and the kaon-to-pion ratios both show a distinct
peak at $\pt \approx \gevc{3}$ in central \pbpb collisions. Below the peak,
$\pt < \gevc{3}$, both ratios are in good agreement with hydrodynamical
calculations, suggesting that the peak itself is dominantly the result of
radial flow rather than anomalous hadronization processes. For $\pt >
\gevc{10}$ particle ratios in \pp and \pbpb collisions are in agreement and
the nuclear modification factors for $\pi^{\pm}$, $\rm K^{\pm}$ and $\rm
p$($\bar{\rm p}$) indicate that, within the systematic and statistical
uncertainties, the suppression is the same. This suggests that the chemical
composition of leading particles from jets in the medium is similar to that of
vacuum jets.
\end{abstract}
\end{titlepage}
\setcounter{page}{2}
%
%

\section{Introduction}

Heavy-ion collisions at ultra relativistic energies produce a new form of QCD
matter characterized by the deconfined state of quarks and gluons
(partons). Measurements of the production of identified particles in \pbpb
collisions, relative to \pp collisions, provide information about the dynamics
of this dense matter. In \pp collisions, high transverse momentum ($\pt >
\gevc{2}$) hadrons are produced from fragmentation of jets that can be
calculated folding the perturbative QCD calculations for jets with universal
fragmentation functions determined from data such as those reported here. The
bulk production of particles at lower \pt is non-perturbative and requires
phenomenological modeling. In heavy-ion collisions the production can be
affected by the medium in several different ways. In particular there is an
intermediate transverse momentum regime, $2 < \pt < \gevc{8}$, where the
baryon-to-meson ratios, {\it e.g.}  the proton yield divided by the pion
yield, measured by experiments at RHIC revealed a, so far, not well understood
enhancement~\cite{Adcox:2001mf,Adler:2003kg,Adams:2003am}. This so-called
``baryon anomaly'' could indicate the presence of new hadronization mechanisms
such as parton
recombination~\cite{PhysRevC.68.044902,Pop:2004dq,Brodsky:2008qp} that could
be significantly enhanced and/or extended out to higher \pt at LHC due to
larger mini-jet production~\cite{Hwa:2006zq}. For transverse momenta above
\gevc{10} one expects to be able to study the pure energy loss (jet quenching)
of high \pt scattered partons traversing the
medium~\cite{Gyulassy:1990ye,Gyulassy:1993hr,Wang:2002ri}.  This affects the
inclusive charged particle \pt spectrum as has been seen at
RHIC~\cite{Adcox:2004mh,Adams:2005dq} and over an extended \pt range, up to
\gevc{100}, at the LHC~\cite{Aamodt:2010jd,CMS:2012aa}. The additional
information provided by particle identification (PID) is of fundamental
interest to study the differences in the dynamics of fragmentation between
quarks and gluons to baryons and mesons~\cite{Abreu:2000nw}, and also to study
the differences in their interaction with the medium considering that, due to
the color Casimir factor, gluons lose a factor of two more energy than
quarks~\cite{Wang:1998bha,Renk:2007rp}. The results presented in this Letter
address three open experimental questions: Are there indications that the
kaons are affected by radial flow at intermediate \pt? Does the
baryon-to-meson ratio return to the \pp value for high \pt ($> \gevc{10}$) as
suggested by the recent publication of the $\Lambda/\rm K^{0}_{S}$
ratio~\cite{Abelev:2013xaa}?  Are there large particle species dependent jet
quenching effects as predicted in several
models~\cite{Sapeta:2007ad,Aurenche:2011rd,Bellwied:2010pr}, where
measurements at RHIC, in particular for baryons, are inconclusive due to the
limited \pt-range and the large systematic and statistical
uncertainties~\cite{Abelev:2006jr,Agakishiev:2011dc,Adare:2013esx}?

\section{Data analyses}

In this Letter we present the measurement of the production of pions (kaons
and protons) from a \pt of a few hundred MeV/$c$ up to $\pt = \gevc{20}$ in
$\sqrt{s_{\rm NN}} =$ 2.76\,TeV \pp and \pbpb collisions with the ALICE
detector~\cite{Aamodt:2008zz}. The Inner Tracking System (ITS) and the Time
Projection Chamber (TPC) are used for vertex finding and tracking. The ITS and
TPC also provide PID through the measurement of the specific energy loss,
\dedx. The PID is further improved at low and intermediate \pt using the
Time-of-Flight (TOF) and the High Momentum PID (HMPID) Cherenkov detectors. In
\pbpb collisions the spectra at low \pt have already been
published~\cite{Abelev:2013vea} and the new addition here is the extension of
the \pt range up to \gevc{20} and the improvement at intermediate \pt for the
0--40\% most central collisions using the HMPID. The \pp low \pt analysis
combining information from ITS, TPC, and TOF follows the same procedures as
the ones published by ALICE at $\sqrt{s} = 900$\,GeV~\cite{Aamodt:2011zj} and
in $\sqrt{s_{\rm NN}} =$ 2.76\,TeV \pbpb collisions~\cite{Abelev:2013vea}. The
main focus in the following will therefore be on explaining the analysis
details for the HMPID and the high \pt \dedx analysis.  \\ The \pp analyses
use 40$\times 10^{6}$ and the \pbpb minimum bias analysis uses 11 $\times
10^{6}$ collision events. The HMPID analysis used the 2011 centrality
triggered \pbpb data with around $4.1 \times 10^6$ 0-5\% central collision
events. Data were taken during 2010 and 2011 under conditions where pileup
effects were negligible. Minimum bias interactions are triggered based on the
signals from forward scintillators (V0) and, in \pp collisions, the two
innermost silicon pixel layers of the ITS (SPD). The trigger efficiency is
88.1\% for \pp inelastic collisions~\cite{Abelev:2012sea} and 97.1\% for
non-diffractive \pbpb collisions~\cite{Abelev:2013qoq}.  The \pbpb collision
centrality is determined from the measured amplitude in the V0
detector~\cite{Abbas:2013taa} which is related to the number of participating
nucleons and the nuclear overlap function ($T_{\rm AA}$) through simulations
based on a Glauber model~\cite{Abelev:2013qoq}. The same event and track
selection is used as in the inclusive charged particle
analysis~\cite{Abelev:2012hxa}. Track cuts are optimized in order to select
primary charged particles in the pseudorapidity range $|\eta| < 0.8$ and all
results presented in this paper are corrected for feed-down from weak
decays. As listed in Table~\ref{tab:ranges_PID} the low \pt analysis is done
for $|y| < 0.5$, while the high \pt analysis is done for $|\eta| < 0.8$, to
take advantage of the full statistics, and the final spectra are then
normalized to the corresponding rapidity intervals, see Eq.~\ref{eq:spectra}
below.

\subsection{Identified particle spectra at low \pt}

The \pp low \pt analysis relies on the combination of four almost independent
PID techniques, named after the detectors involved: ITS-sa, TPC-TOF, TOF and
HMPID. The techniques have complementary \pt ranges listed in
Table~\ref{tab:ranges_PID}.
\begin{table}[tp]
  \centering
  \begin{tabular}{ccccc}
    \hline \hline
    Analysis & $\eta/y$ range & $\pi$ & K & p \\
    \hline
    ITS-sa    & $|y| < 0.5$ & 0.1-0.7  & 0.2-0.55  & 0.3-0.6   \\
    TPC--TOF  & $|y| < 0.5$ & 0.3-1.2 & 0.3-1.2  & 0.45-2.0  \\
    TOF       & $|y| < 0.5$ & 0.5-2.5  & 0.5-2.4  & 0.8-3.8   \\
    HMPID     & $|y| < 0.5$ & 1.5-4.0  & 1.5-4.0  & 1.5-6.0   \\
    High \pt \dedx  &  $|\eta| < 0.8$ & 2.0-20.0  & 3.0-20.0  & 3.0-20.0   \\
    \hline \hline
  \end{tabular}
  \caption{\label{tab:ranges_PID} The $\eta/y$ and \pt range (GeV/$c$) covered
    by each analysis.}
\end{table}

The ITS-sa analysis exploits stand-alone (sa) tracks reconstructed in the ITS
to be able to go as low in \pt as possible. The identification is done based
on \dedx measurements in up to 4 of the 6 silicon layers. This information is
combined in a Bayesian approach using a set of priors determined with an
iterative procedure, and the track identity is assigned according to the
highest probability. The minor residual contamination due to misidentification
is less than $10\%$ in the \pt-range reported in Table~\ref{tab:ranges_PID}
and corrected for using MC.

The other three analyses all use global tracks reconstructed in both the ITS
and the TPC. The TPC-TOF analysis is optimized to combine the information from
the TPC and TOF. The identification is based on a three standard deviations
agreement with the expected detector signal and resolution ($3\sigma$) in the
TPC \dedx and for $\pt > \gevc{0.6}$ a $3\sigma$ requirement is also applied
for the time-of-flight provided by the TOF detector. The TOF analysis
identifies particles comparing the measured time-of-flight from the primary
vertex to the TOF detector, $t_{\rm{tof}}$, and the time expected under a
given mass hypothesis, $t^{exp}_{i}$ ($i$ = $\pi$, K, p). The TOF standalone
analysis is optimized for handling momentum regions where the separation is
challenging. The precise signal shape for $t_{\rm{tof}} - t^{exp}_{i}$,
including an exponential tail, is used, and the yield in a given \pt interval
is obtained by fitting.

\begin{figure}[h]
  \centering
  \includegraphics[width=0.49\textwidth]{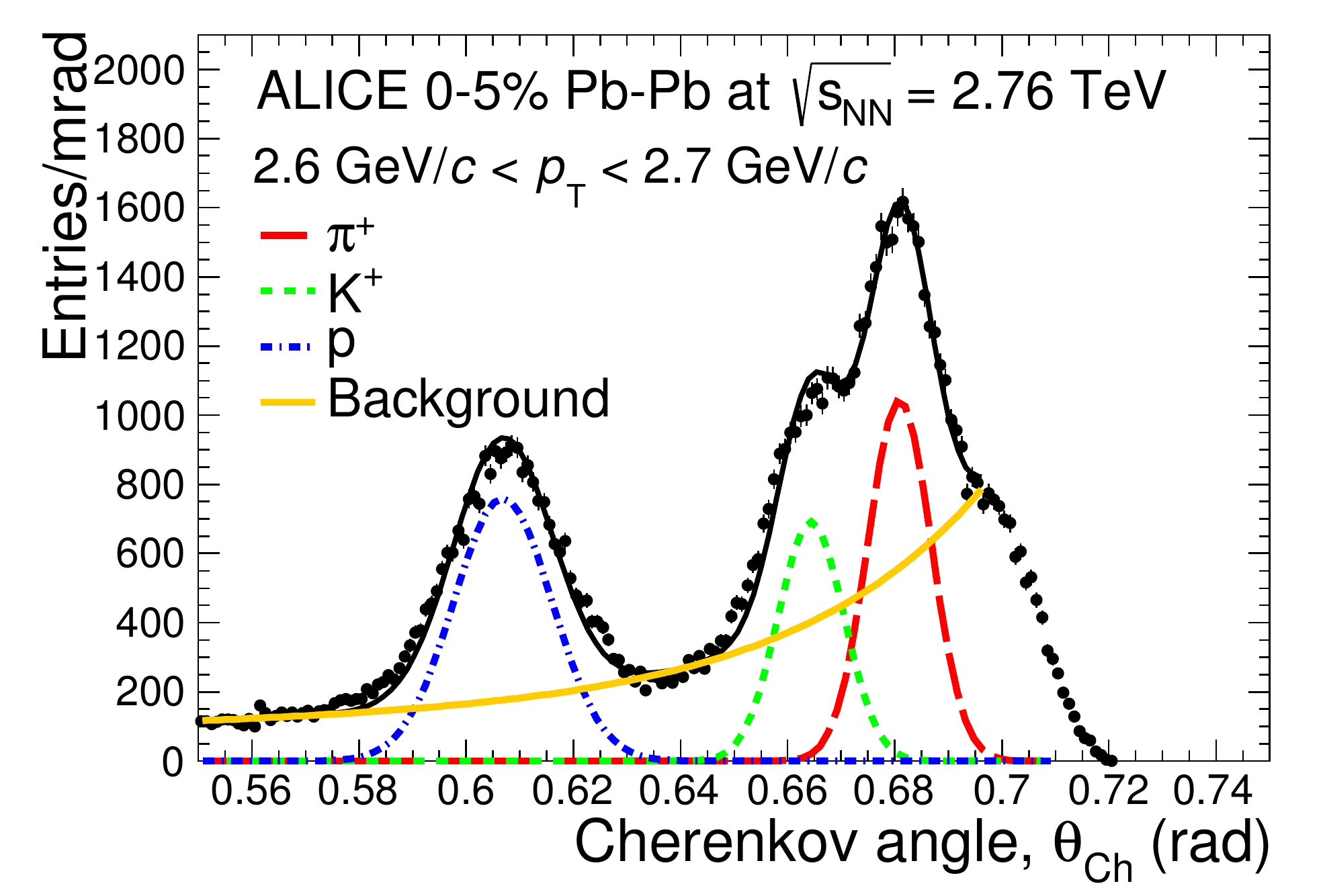}
  \includegraphics[width=0.49\textwidth]{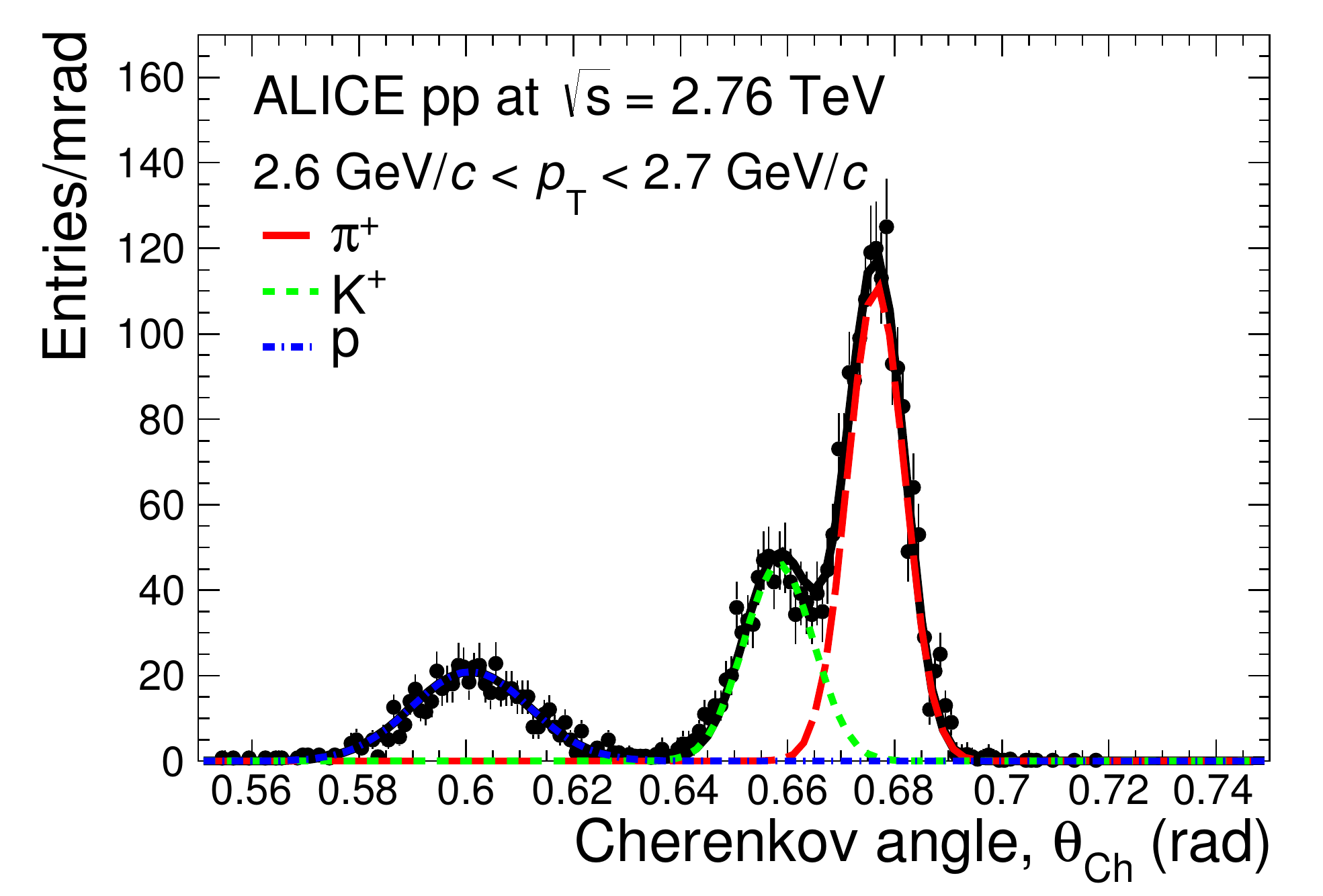}
  \caption{Distributions of the Cherenkov angle measured in the HMPID for
    positive tracks in a narrow \pt bin, for 0-5\% central \pbpb (left) and pp
    (right) collisions.}
  \label{fit}
\end{figure}

The HMPID~\cite{hmpid:tdr,Martinengo:2011zza} is designed as a single-arm
proximity-focusing Ring Imaging CHerenkov (RICH) detector where the radiator
is a 15\,mm thick layer of liquid C$_6$F$_{14}$ (perfluorohexane). It is
located at about 5\,m from the beam axis, covering a limited acceptance of
$|\eta| < 0.55$ and $1.2^\circ < \varphi < 58.5^\circ$. The PID in the HMPID
is done by measuring the Cherenkov angle, $\theta_{\text{ch}}$. In the
reconstruction, the tracks are propagated to the HMPID detector and associated
with a MIP signal. A Hough Transform Method (HTM)~\cite{Cozza:2001zz} is used
to discriminate the signal from the background. For a given track, the mean
Cherenkov angle is computed as the weighted average of the single photon
angles selected by the HTM. The Cherenkov angle distribution is then fitted to
obtain the yields, see Fig.~\ref{fit} for an example of fits in the \pbpb and
\pp analysis.

The raw yields measured by each analysis are corrected for the reconstruction,
selection, PID efficiency, and misidentification probability. The
contamination due to particles from weak decays of light flavor hadrons and
interactions with the material is subtracted using MC-template fits of the
distance-of-closest approach distributions~\cite{Abelev:2013vea}. Finally the
raw spectra are corrected for the detector acceptance, trigger selection,
vertex and track reconstruction efficiency.

The systematic uncertainties for the ITS, TPC, and TOF analyses are obtained
in essentially the same way as reported
in~\cite{Aamodt:2011zj,Abelev:2013vea}. The systematic uncertainty for the
HMPID analysis has contributions from tracking and PID. These uncertainties
have been estimated by changing individually the track selection cuts and the
parameters of the fit function used to extract the raw yields by $\pm 10\%$.
In addition, the uncertainty of the association of the track to the MIP signal
is obtained by varying the value of the distance cut required for the match.

The HMPID analysis in \pbpb collisions is analogous to the \pp analysis except
for the treatment of the background. In central \pbpb collisions, where the
total number of hits in the HMPID chambers is large, it is possible that a
Cherenkov ring is constructed based on hits incorrectly associated with the
track. Figure~\ref{fit} gives examples of the reconstructed Cherenkov angle
distributions in a narrow \pt interval. In \pp collisions (right panel) the
reconstructed angle distribution is fitted by a sum of three Gaussian
distributions, corresponding to the signals from pions, kaons and protons.  In
the case of \pbpb collisions (left panel) the additional background
distribution is modeled with a $6^{\text{th}}$ order polynomial found to
minimize the reduced $\chi^{2}$ of the fit. The shoulder in the background
distribution starting at 0.7 rad is a boundary effect due to the finite
chamber geometrical acceptance that is also observed in MC simulations. The
fitting is done in two steps, where the width and the mean of each Gaussian
distribution are free parameters in the first step and are then used to obtain
a \pt dependent parameterization. This parameterization is used to constrain
the parameters in the second final fit. The means and widths constrained in
this way are found to be independent of centrality. Finally we note that the
background increases with the Cherenkov angle because the fiducial area used
in the reconstruction becomes larger, making it more likely to associate
spurious hits with the signal.

The PID efficiency has been evaluated from a Monte Carlo simulation that
reproduces well the background in the data. A data-driven cross check of the
efficiency has been performed using a clean sample of protons and pions from
$\Lambda$ and \ks decays identified in the TPC based on their topological
decay.

To estimate the uncertainty due to the incomplete knowledge of the shape of
the background distribution, an alternative background function, depending on
$\tan(\theta_{\text{ch}})$ and derived from geometrical considerations in case
of orthogonal tracks~\cite{hmpid:tdr}, has been used. The corresponding
systematic uncertainty reaches the maximum value at low momenta for the most
central collisions (${\sim}15\%$ for pions, and ${\sim}8\%$ for kaons
and protons). The systematic uncertainty decreases with \pt as the track
inclination angle in the bending plane decreases so that the fiducial area for
the Cherenkov pattern search is smaller.

\subsection{Identified particle spectra at high \pt}

 \begin{figure}[htbp]
\includegraphics[keepaspectratio, width=1.0\columnwidth]{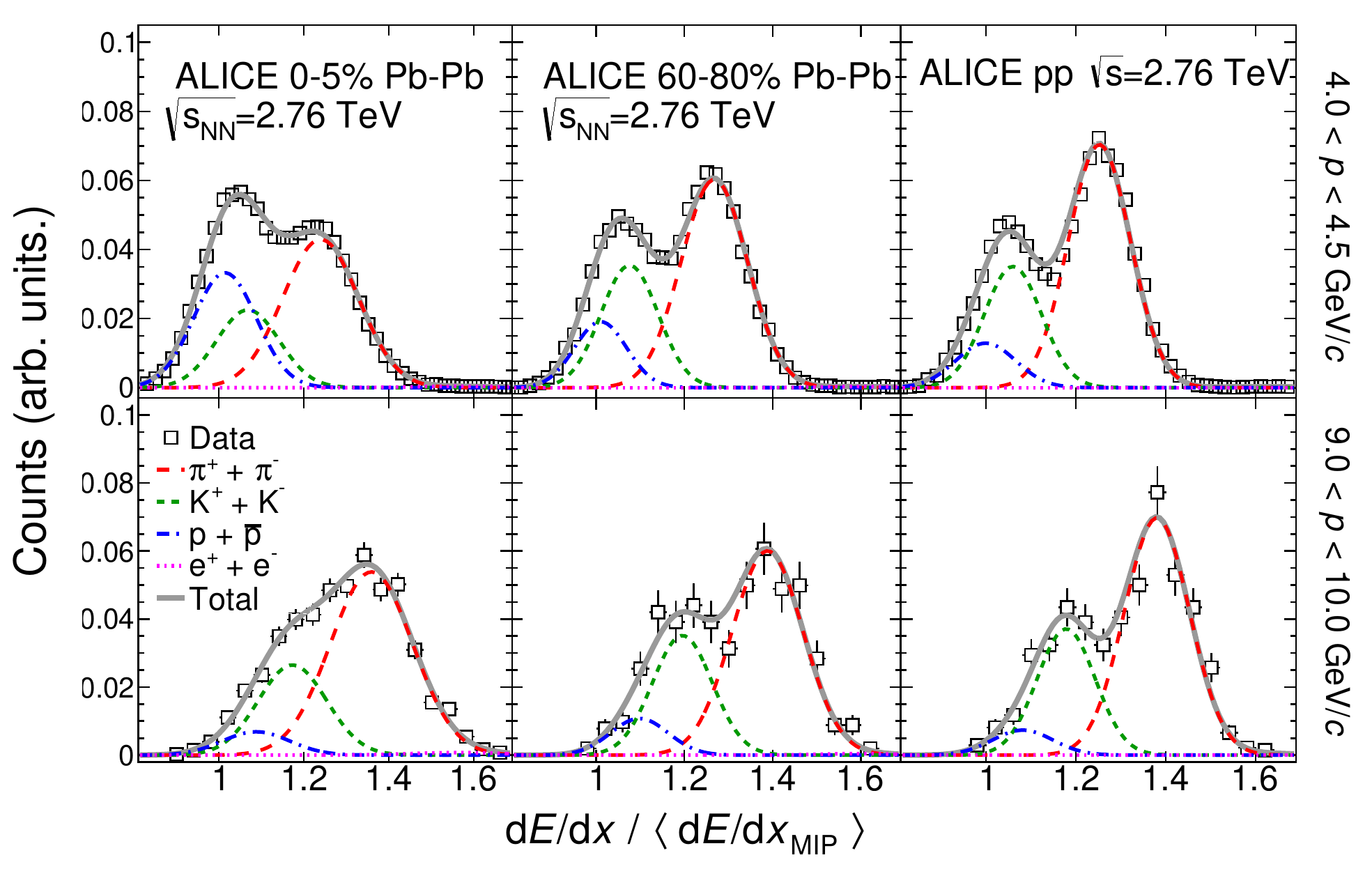}
\caption{\label{fig:1} \dedx distributions measured for $|\eta| < 0.2$ and
  normalized to the integrated yields. The signals are fitted to a sum of four
  Gaussian functions (solid line). Two $p$ intervals are shown for central
  (left) and peripheral (center) \pbpb; and \pp (right) collisions. In all
  momentum intervals the electron fraction is below 1\% (not
  visible). Individual yields are shown as dashed curves; protons in blue
  (left), kaons in green, and pions in red (right).}
\end{figure}

Particle identification is performed in the relativistic rise regime of the
Bethe-Bloch (BB) curve where the \mdedx separation between particles with
different masses is nearly constant~\cite{Alessandro:2006yt}. The \dedx is
obtained as the truncated mean of the 0-60\% lowest charge samples associated
with the track in the TPC~\cite{Alme:2010ke}. The \dedx response depends on
the track length so the analysis is done in four equally sized
$|\eta|$-intervals, and a geometrical cut to remove tracks entering the gap in
between the TPC readout chambers is applied to select tracks with the best
\dedx resolution. The separation in number of standard deviations ($\sigma$)
between pions and kaons (pions and protons) in \pp and peripheral \pbpb
collision is around 3.2 (4.6) at momentum $p \approx \gevc{6}$ for $0.6 <
|\eta| < 0.8$ where the separation is largest. In central \pbpb collisions one
finds a separation of 2.4$\sigma$\,(3.5$\sigma$). In the worst case, $|\eta| <
0.2$, the separation is 11-15\% smaller.

Figure~\ref{fig:1} shows examples of the \dedx spectra obtained for \pp and
\pbpb (central and peripheral) collisions for two momentum, $p$, intervals and
$|\eta| < 0.2$ where $p \approx \pt$. The pion, kaon, and proton yields are
extracted by fitting a sum of four Gaussian functions (including electrons) to
the \dedx spectra~\footnote{We note that muons from heavy flavor decays are
  subtracted from the pions based on the measured electron yields and that
  contamination from deuterons and tritons are negligible ($\ll 1\%$).}. To
reduce the degrees of freedom in the fits from 12 to 4, parameterizations of
the BB (\mdedx) and resolution ($\sigma$) curves as a function of \bg are
extracted first using tracks from identified particles. Samples of secondary
pions ($30 < \bg < 50$) and protons ($3 < \bg < 7$) are obtained through the
reconstruction of the weak-decay topology of \ks and \lmd, respectively; a
similar algorithm is used to identify electrons resulting from photon
conversions (fixing the \dedx plateau: $\bg > 1000$) . Finally, using
information from the time-of-flight detector the relative pion content can be
enhanced for sub-samples of the full datasets ($16 < \bg < 50$).

The \mdedx separation between kaons and protons in the high \pt analysis is
smallest for $p \approx \gevc{3}$ and increases with $p$ until both species
are on the relativistic rise~\cite{Alessandro:2006yt}. In central collisions
the \mdedx separation is the lowest and the systematic uncertainties on the
extracted yields are correspondingly large as discussed later, see
table~\ref{tab:table1}. Hence, to improve the central values for the kaons and
protons, the \ks yields~\cite{Abelev:2013xaa} are used as a proxy for the
charged kaons to further constrain the BB curve in \pbpb collisions in a
procedure which uses a two dimensional fit of \dedx vs momentum. The effect of
the ${\rm K_{S}^{0}}$ bias is only relevant in central collisions at low \pt
($< \gevc{4}$). At \gevc{3} the effect on the extracted kaon yield is an
increase of 10\% ($<1$\%) for 0-5\% (60-80\%) collision centrality.

With the above information the BB and the resolution curves are determined for
kaons and protons in the full momentum interval reported here and for pions
with $p < \gevc{7}$. For $p > \gevc{7}$ the pion \mdedx is restricted by the
logarithmic rise until the \mdedx starts to approach the plateau. This lack of
additional constraint currently limits the \pt reach of the analysis to
${\sim}\gevc{20}$.

From the fits in Fig.~\ref{fig:1} the particle fractions, $f_{\pi/{\rm
    K/p}}(p)$ are extracted. The fraction in a \pt bin, $f_{\pi/{\rm
    K/p}}(\pt)$, is obtained as the weighted average of the contributing
momentum ($p$) bins. The \pt-dependent fractions are found to be independent
of $\eta$ and so all four $\eta$ regions are averaged.

Finally, the invariant yields are obtained using the \pt spectrum for inclusive charged particles~\cite{Abelev:2012hxa}, $\frac{{\rm d}^{2}N_{\rm ch}}{{\rm d}p_{\rm T}{\rm d}\eta}$, in the following way:
\begin{equation}
\label{eq:spectra}
\frac{{\rm d}^{2}N_{\pi/{\rm K/p}}}{{\rm d}p_{\rm T}{\rm d}y} = {\rm J}_{\pi/{\rm K/p}}\frac{{\rm d}^{2}N_{\rm ch}}{{\rm d}p_{\rm T}{\rm d}\eta} \frac{\epsilon_{\rm ch}}{\epsilon_{\pi/{\rm K/p}}} f_{\pi/{\rm K/p}}(\pt),
\end{equation}
where ($\epsilon_{\rm ch}$) $\epsilon_{\pi/{\rm K/p}}$ is the efficiency for
(un)identified particles and ${\rm J}_{\pi/{\rm K/p}}$ is the Jacobian
correction (from $\eta$ to $y$). Normalizing to the \pt spectrum of inclusive
charged particles guarantees that only the systematic uncertainty due to PID
is relevant when comparing the modification of the \pt spectra of $\pi{\rm
  /K/p}$ to those for the unidentified particles. The \pt resolution is around
5\% at $\pt = \gevc{20}$ and the \pt spectra have been corrected for this
resolution using an unfolding procedure for $\pt >
\gevc{10}$~\cite{Abelev:2012hxa,Abelev:2013ala}. This correction is less than
2\% at $\pt = \gevc{20}$.

\subsubsection{Systematic uncertainties}

\begin{table}[ht]
  \centering
  \begin{threeparttable}[c]
    \begin{tabular}{lccc}
      \hline \hline \textrm{System}& \textrm{\pbpb 0-5\%}& \textrm{\pbpb 60-80\%}&
      \textrm{\pp}\\ \hline $N_{\rm ch}$\tnote{a} & 8.3-8.2\% & 9.9-9.8\% & 7.4-7.6\%
      \\ $\pi^{+}+\pi^{-}$\tnote{b} & 1.7-2.4\% & 1.5-2.2\% & 1.2-1.7\% \\ ${\rm
        K}^{+}+{\rm K}^{-}$ & 19-7.9\% & 17-8.7\% & 16-5.7\% \\ ${\rm p}+{\rm
        \bar{p}}$ & 9.9-21\% & 20-24\% & 24-20\% \\ \hline Efficiency
      ratios\tnote{c} & & 3\% &
      \\ \hline \hline
    \end{tabular}
    \begin{tablenotes}[para]
    {\scriptsize
    \item [a] Taken directly from~\cite{Abelev:2012hxa}. \item [b] Additional contribution due to $\mu^{\pm}$ contamination is
      $\leq$1\%.
    \item[c] Same for all centralities and all particle species.
      }
    \end{tablenotes}
  \end{threeparttable}
    \caption{\label{tab:table1} Systematic uncertainties, separated into the
      $N_{\rm ch}$, PID, and efficiency part, on the invariant yields from $3
      < \pt < \gevc{4}$ (left quoted value) to $10 < \pt < \gevc{20}$ (right
      quoted value).}
\end{table}

The systematic uncertainty on the invariant yields has three main components:
event and track selection, efficiency correction of the fractions, and the
fraction extraction. Contributions from the event and track selection are
taken directly from the inclusive charged particle
analysis~\cite{Abelev:2012hxa}. Efficiency ratios ($\epsilon_{\rm
  ch}/\epsilon_{\pi/{\rm K/p}}$) are found to be nearly independent of \pt (a
small dependence is only observed for kaons), similar for all systems, and
model independent within 3\%. The largest systematic uncertainty in the
extraction of the fractions comes from the uncertainty in the constrained
parameters: the means ( \mdedx) and the widths ($\sigma$) used in the
fits. The uncertainty on these parameters are estimated from the average
difference between the final parameterizations and the data points obtained
from the enhanced samples with identified particles. In addition, the
statistical uncertainty on the extracted BB parameterization in peripheral
\pbpb collisions is found to be of a similar magnitude and also taken into
account in the following variations. The \dedx spectra are then refitted,
varying the means and the widths within the estimated uncertainties, and the
variation of the fractions are assigned as systematic errors. In this way the
correlations in the systematic uncertainty for the particle ratios can be
directly included. A summary of the PID systematic uncertainties is shown in
Table~\ref{tab:table1}. The $N_{\rm ch}$ systematic uncertainties cancel in
the particle ratios.

\section{Results and discussion}

\begin{figure*}[htbp]
\centering
\includegraphics[keepaspectratio, width=0.7\columnwidth]{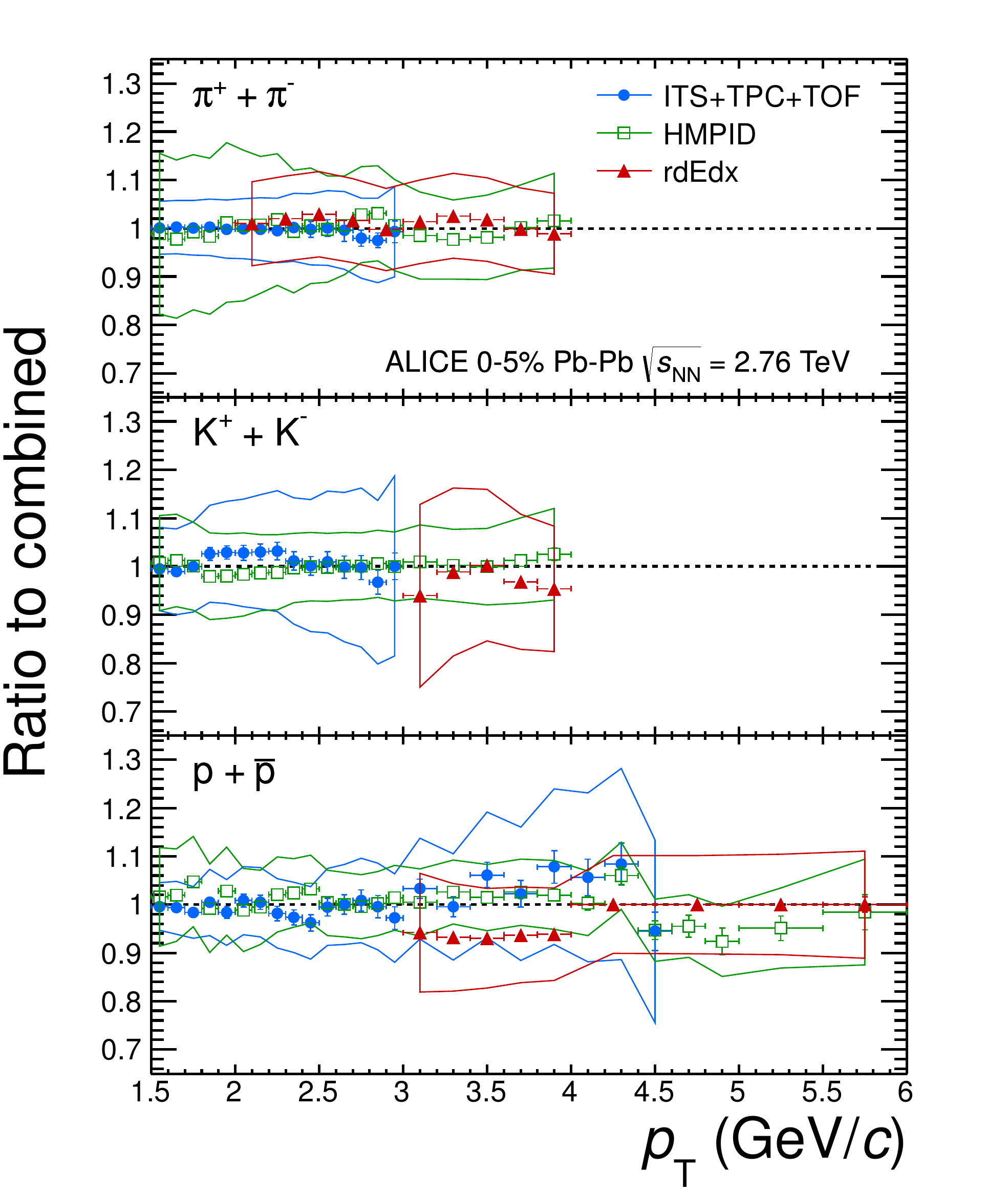}
\caption{\label{fig:combination} The ratio of individual spectra to the
  combined spectrum as a function of \pt for pions (top), kaons (center), and
  protons (bottom). Only the \pt-range where the analyses overlap is
  shown. The ITS+TPC+TOF spectra are the results published
  in~\cite{Abelev:2013vea}. The statistical and independent systematic
  uncertainties are shown as vertical error bars and as a band, respectively,
  and only include those on the individual spectra.}
\end{figure*}

The measurement of charged pion, kaon and (anti-)proton transverse momentum
spectra has been performed via several independent analyses, each one focusing
on a sub-range of the total \pt distribution, with emphasis on the individual
detectors and specific techniques to optimize the signal extraction. The
results were combined using the independent systematic uncertainties as
weights in the overlapping ranges (a $3\%$ common systematic uncertainty due
to the TPC tracking is not in the weight but added directly to the combined
spectrum). The statistical uncertainties are much smaller and therefore
neglected in the combination weights. For $\pt > \gevc{4}$ only the high \pt
analysis is used for all species. Figure~\ref{fig:combination} shows the ratio
of individual spectra to the combined spectrum for the 0-5\% central \pbpb
data, illustrating the compatibility between the different analyses.

\begin{figure*}[htbp]
\includegraphics[keepaspectratio, width=1.0\columnwidth]{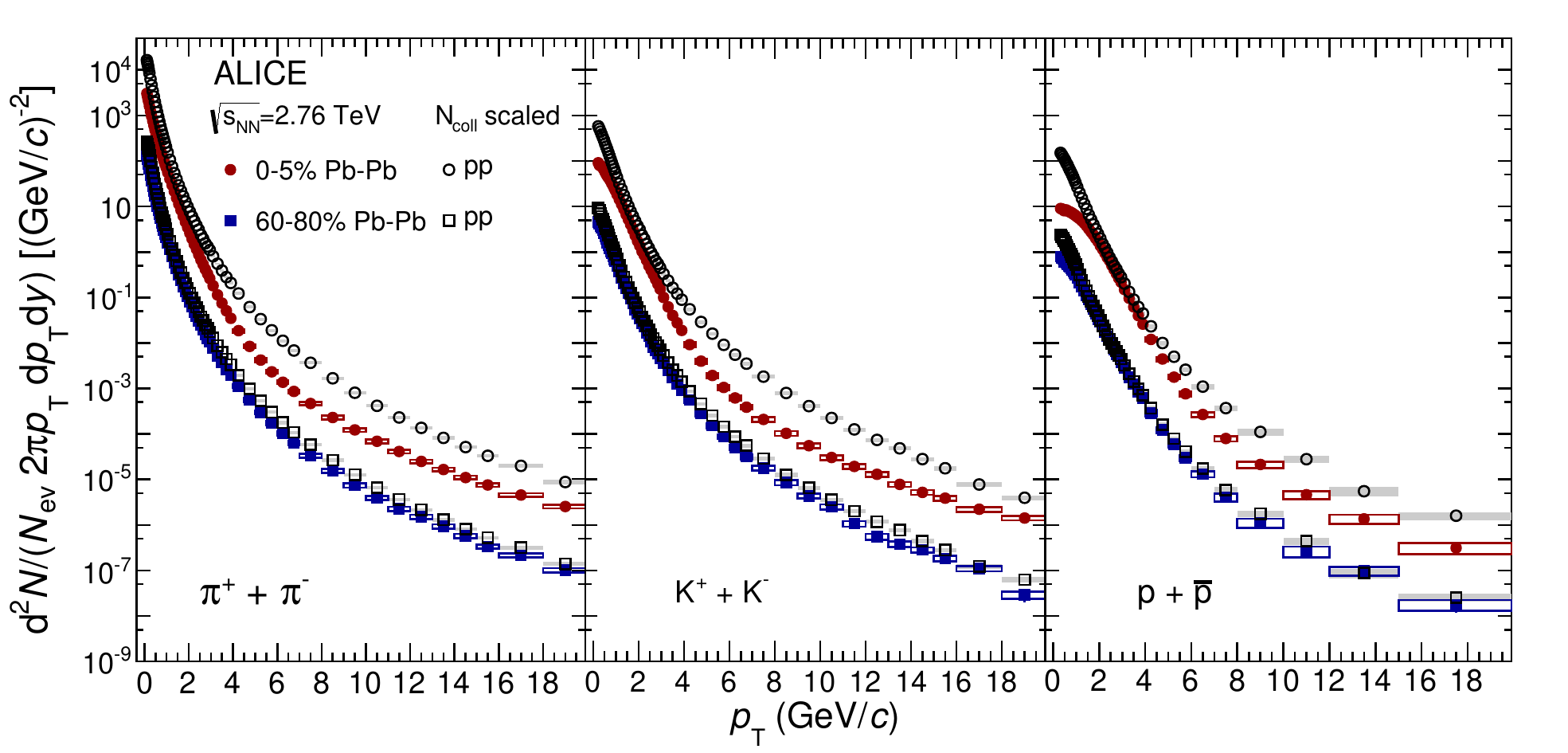}
\caption{\label{fig:2} Solid markers show the invariant yields of identified
  particles in central (circles) and peripheral (squares) \pbpb
  collisions. Open points show the \pp reference yields scaled by the average
  number of binary collisions for 0-5\% (circles) and 60-80\%
  (squares)~\cite{Abelev:2013qoq}. The statistical and systematic
  uncertainties are shown as vertical error bars and boxes, respectively.}
\end{figure*}

\begin{figure*}[htbp]
\includegraphics[keepaspectratio, width=1.0\columnwidth]{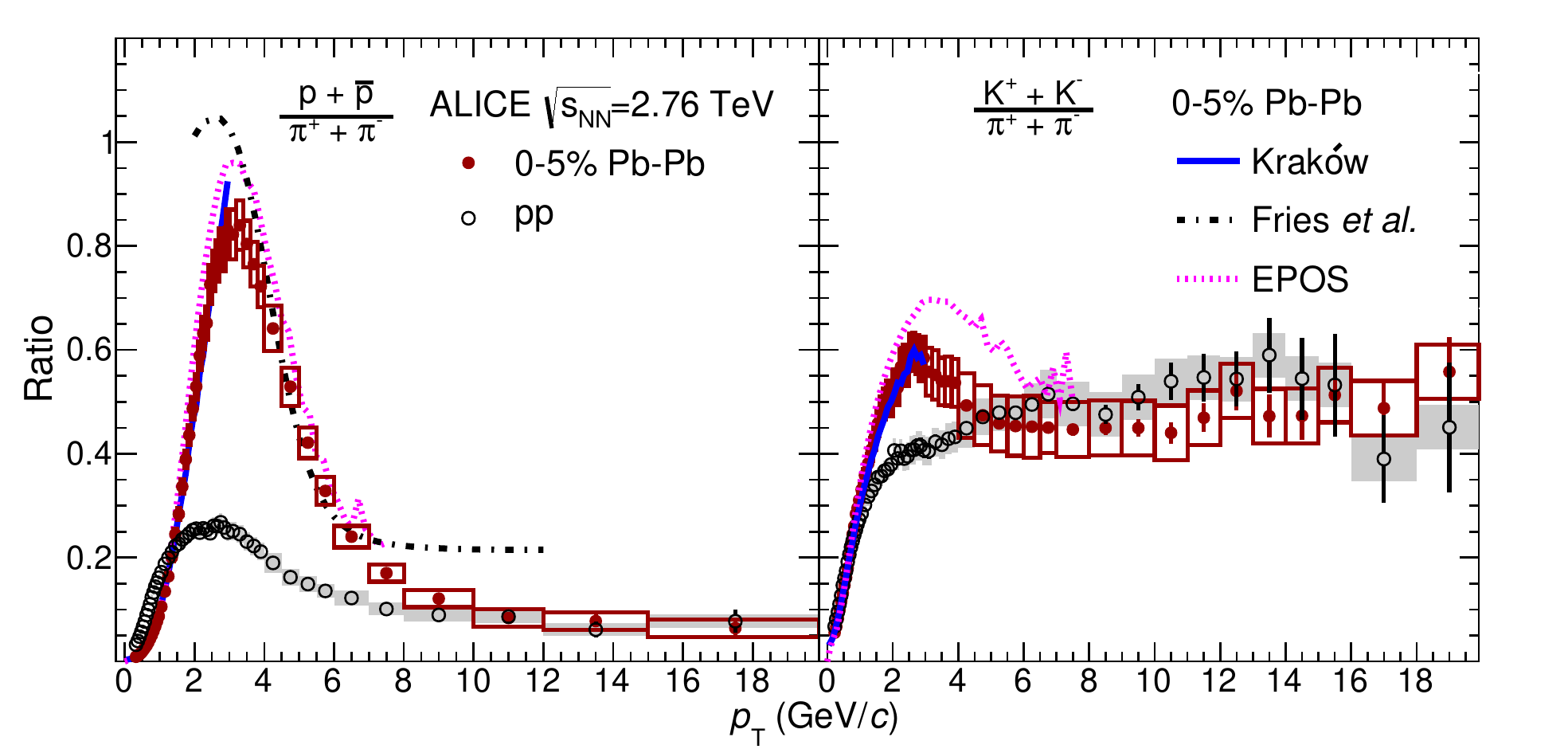}
\caption{\label{fig:3} Particle ratios as a function of \pt measured in \pp
  and the most central, 0-5\%, \pbpb collisions. Statistical and systematic
  uncertainties are displayed as vertical error bars and boxes,
  respectively. The theoretical predictions refer to \pbpb collisions, see
  text for references.}
\end{figure*}

Figure~\ref{fig:2} shows the invariant yields measured in \pbpb collisions
compared to those in \pp collisions scaled by the number of binary collisions,
$N_{\rm coll}$~\cite{Abelev:2013qoq} obtained for the measured \pp cross
section~\cite{Abelev:2012sea}. For peripheral \pbpb collisions the
shapes of the invariant yields are similar to those observed in \pp
collisions. For central \pbpb collisions, the spectra exhibit a
reduction in the production of high-\pt particles with respect to the
reference which is characteristic of jet quenching.

Figure~\ref{fig:3} shows the proton-to-pion ratio, $({\rm
  p+\bar{p}})/(\pi^{+}+\pi^{-})$, as a function of \pt. For central
(peripheral, not shown) \pbpb collisions it reaches ${\sim}0.83$ (${\sim}0.35$)
at the maximum around \gevc{3} and then decreases with increasing \pt. These
values are approximately 20\% above the peak values measured by
PHENIX~\cite{Adare:2013esx} and STAR~\cite{Abelev:2006jr}, when ${\rm
  p}/\pi^+$ and ${\rm \bar{p}}/\pi^-$ ratios are averaged and data are
corrected for feed-down.

At LHC energies the mini-jet activity is expected to be larger than at RHIC
energies, which motivated ratio predictions in the framework of recombination
models where shower partons in neighboring jets can recombine to be an order
of magnitude larger than the measurements reported
here~\cite{Hwa:2006zq}. Other predictions where recombination only occurs for
soft thermal radially flowing partons are, as shown in the figure, more
consistent with the data~\cite{PhysRevC.68.044902}. The surprising new result
is that in central \pbpb collisions the \ktopi ratio also exhibits a bump at
$\pt \approx \gevc{3}$. This has not been observed at RHIC (this could be due to
limitations in precision in this \pt region) but is also observed in the soft
coalescence model~\cite{PhysRevC.68.044902}. The
Krak{\'o}w~\cite{Bozek:2012qs} hydrodynamical model captures the rise of both
ratios quantitatively well, while a similar model, HKM~\cite{Karpenko:2012yf}
that is not shown, does slightly worse. The EPOS~\cite{PhysRevC.85.064907}
event generator which has both hydrodynamics, but also the high \pt physics
and special hadronization processes for quenched jets ~\cite{Werner:2012sv}
qualitatively well describes the data but tends to overestimate the peaks. The
recent result~\cite{Abelev:2014uua} that for $\pt < \gevc{3}$ the shape of the
phi-to-pion ratio is consistent with the proton-to-pion ratio, reported here,
taken together with the model comparisons shown in Fig.~\ref{fig:3} indicate
that the peak is mainly dominated by radial flow (the masses of the hadrons).

For higher \pt ($> \gevc{10}$) both particle ratios behave like those in \pp,
suggesting that fragmentation dominates the hadron production. In this \pt
regime, the particle ratios in \pp are not well described by the pQCD
calculations in~\cite{PhysRevD.82.074011}. It was recently
shown~\cite{d'Enterria:2013vba} that in general the fragmentation functions
for gluons are badly constrained, leading to disagreement of up to a factor 2
with $N_{\rm ch}$ spectra measured at LHC. Furthermore it was pointed out that
data with $\pt > \gevc{10}$, as reported here, are needed to reduce the scale
dependence that seems to be the origin of the disagreement.\\

\begin{figure*}[htbp]
\includegraphics[keepaspectratio, width=1.0\columnwidth]{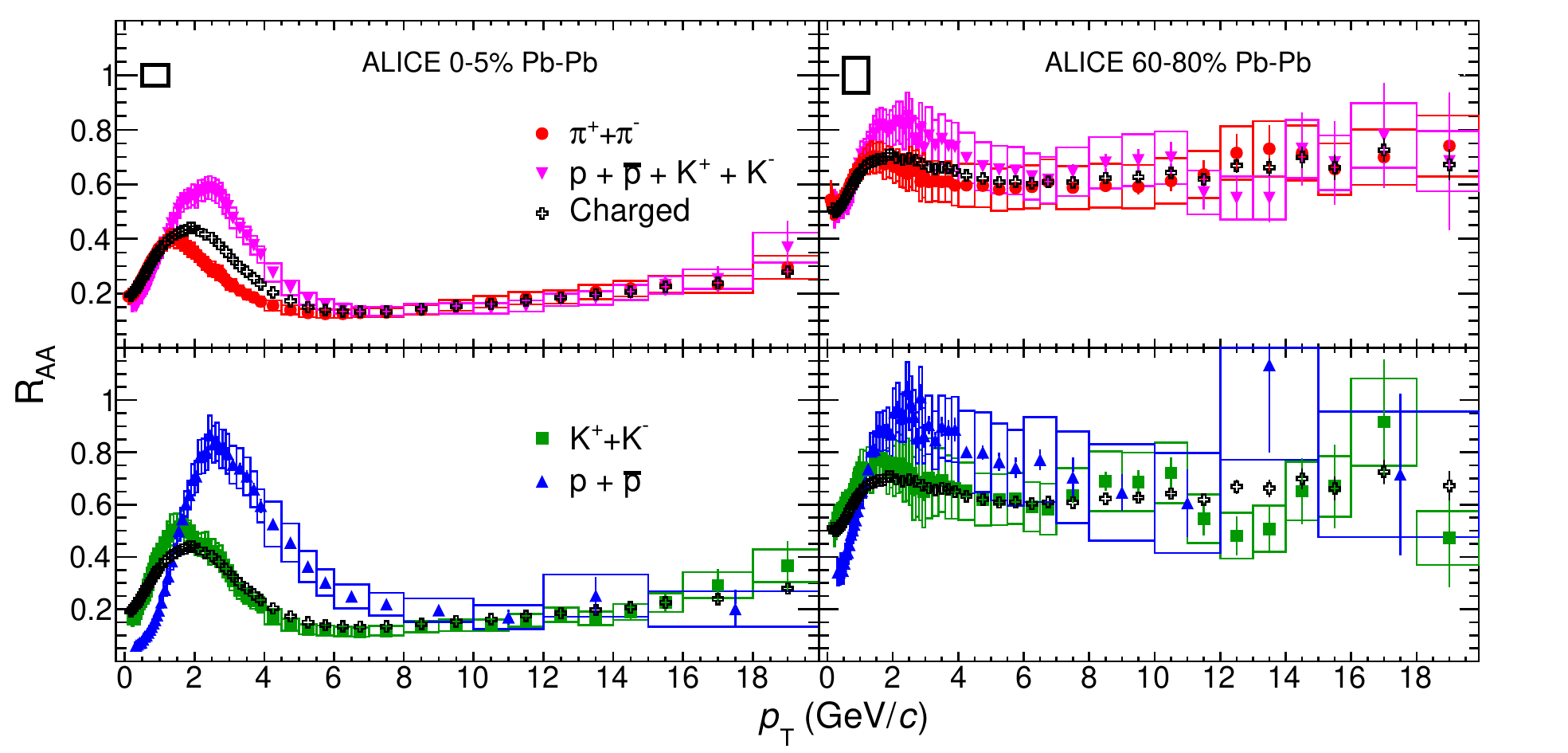}
 \caption{\label{fig:4} The nuclear modification factor \raa as a function of
   \pt for different particle species. Results for 0-5\% (left) and 60-80\%
   (right) collision centralities are shown. Statistical and systematic
   uncertainties are plotted as vertical error bars and boxes around the
   points, respectively. The total normalization uncertainty (\pp and \pbpb)
   is indicated by the black boxes in the top panels~\cite{Abelev:2012hxa}.}
\end{figure*}

Figure~\ref{fig:4} shows the nuclear modification factor \raa as a function of
\pt defined as the ratio of the \pbpb spectra to the $N_{\rm coll}$ scaled \pp
spectra shown in Fig.~\ref{fig:2}. The \raa for the sum of kaons and protons
is included as it allows the most precise quantitative comparison to the \raa
of pions.  For $\pt < \gevc{10}$ protons appear to be less suppressed than
kaons and pions, consistent with the particle ratios shown in
Fig.~\ref{fig:3}. At larger \pt ($> \gevc{10}$) all particle species are
equally suppressed; so despite the strong energy loss observed in the most
central heavy-ion collisions, the particle composition and ratios at high \pt
are similar to those in vacuum.

\begin{figure*}[htbp]
\centering
\includegraphics[keepaspectratio, width=0.7\columnwidth]{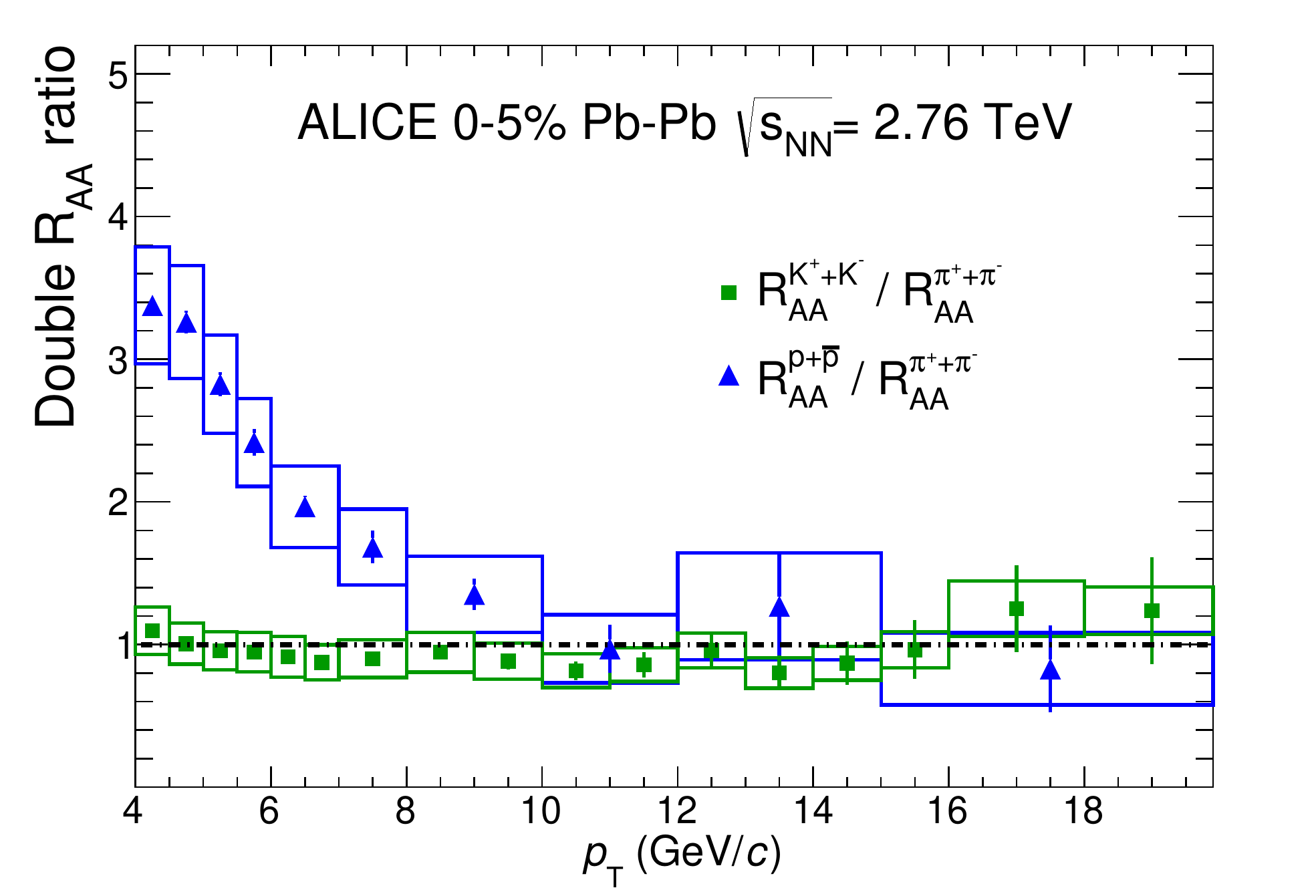}\\
 \caption{\label{fig:5} \raa double ratios as a function of \pt for $\pt >
   \gevc{4}$. Statistical and PID systematic uncertainties are plotted as
   vertical error bars and boxes around the points, respectively.}
\end{figure*}

The models cited in the introduction all suggest large differences, of 50\%
or more, between the suppression of different species that are either related
to mass ordering or baryon-vs-meson effects. The differences are naturally
large in these scenarios because they are directly related to the large
suppression. To quantify the similarity of the suppression the double \raa
ratios, {\it e.g.} $\raa^{{\rm p+\bar{p}}}/\raa^{\pi^{+}+\pi^{-}}$, are
inspected. Figure~\ref{fig:5} shows the double ratios constructed using the
particle ratios to properly handle that the dominant correlated systematic
uncertainties are between particle species and not between different collision
systems. We note that a similar ratio for protons and pions made with the STAR
data~\cite{Abelev:2006jr,Agakishiev:2011dc} would give a flat ratio for $\pt >
\gevc{3}$ of approximately $3 \pm 2$. The results disfavor significant
modifications of hadro-chemistry within the hard core of jets, as predicted
based on medium modified color flow which introduces a mass ordering of the
fragmentation~\cite{Sapeta:2007ad}, or due to changes in the color structure
of the quenched probe which could enhance baryon
production~\cite{Aurenche:2011rd}. The data also contradict predictions where
fragmentation into color neutral hadrons, assumed to have no energy loss after
formation, occurs in the medium and the formation time scales directly with
the hadron mass~\cite{Bellwied:2010pr}.

\section{Conclusions}

The production of pions, kaons and protons has been measured in \pp and
central and peripheral \pbpb collisions up to high \pt. From the invariant
yields we derived the particle ratios and the \raa as a function of \pt. We
observe that the proton-to-pion and the kaon-to-pion ratios both exhibits a
peak and that at low \pt the rise of both ratios can be well described by
hydrodynamic calculations. This rules out models where shower partons
recombine and sets strong constraints for soft recombination models. At
higher-\pt, both ratios are compatible with those measured in \pp collisions.
From the nuclear modification factor \raa, we conclude that for $\pt >
\gevc{10}$ within the systematic and statistical uncertainties, pions, kaons
and protons are suppressed equally. This rules out ideas in which the large
energy loss leading to the suppression is associated with strong mass ordering
or large fragmentation differences between baryons and mesons. The results
presented here establish strong constraints on theoretical modeling for
fragmentation and energy loss mechanisms.

\newenvironment{acknowledgement}{\relax}{\relax}
\begin{acknowledgement}
\section*{Acknowledgements}
\input{acknowledgements_march2013.tex}    
\end{acknowledgement}
%
%
\bibliographystyle{apsrev4-1}   
\bibliography{alicepreprint_CDS.bbl}
%


\newpage
%
%
\appendix
\section{The ALICE Collaboration}
\label{app:collab}
\input{Alice_Authorlist_2013-Dec-03-CERNPREP.tex}  
\end{document}

%% file: acknowledgements_march2013.tex
The ALICE collaboration would like to thank all its engineers and technicians for their invaluable contributions to the construction of the experiment and the CERN accelerator teams for the outstanding performance of the LHC complex.
\\
The ALICE collaboration gratefully acknowledges the resources and support provided by all Grid centres and the Worldwide LHC Computing Grid (WLCG) collaboration.
\\
The ALICE collaboration acknowledges the following funding agencies for their support in building and
running the ALICE detector:
 \\
State Committee of Science,  World Federation of Scientists (WFS)
and Swiss Fonds Kidagan, Armenia,
 \\
Conselho Nacional de Desenvolvimento Cient\'{\i}fico e Tecnol\'{o}gico (CNPq), Financiadora de Estudos e Projetos (FINEP),
Funda\c{c}\~{a}o de Amparo \`{a} Pesquisa do Estado de S\~{a}o Paulo (FAPESP);
 \\
National Natural Science Foundation of China (NSFC), the Chinese Ministry of Education (CMOE)
and the Ministry of Science and Technology of China (MSTC);
 \\
Ministry of Education and Youth of the Czech Republic;
 \\
Danish Natural Science Research Council, the Carlsberg Foundation and the Danish National Research Foundation;
 \\
The European Research Council under the European Community's Seventh Framework Programme;
 \\
Helsinki Institute of Physics and the Academy of Finland;
 \\
French CNRS-IN2P3, the `Region Pays de Loire', `Region Alsace', `Region Auvergne' and CEA, France;
 \\
German BMBF and the Helmholtz Association;
\\
General Secretariat for Research and Technology, Ministry of
Development, Greece;
\\
Hungarian OTKA and National Office for Research and Technology (NKTH);
 \\
Department of Atomic Energy and Department of Science and Technology of the Government of India;
 \\
Istituto Nazionale di Fisica Nucleare (INFN) and Centro Fermi -
Museo Storico della Fisica e Centro Studi e Ricerche "Enrico
Fermi", Italy;
 \\
MEXT Grant-in-Aid for Specially Promoted Research, Ja\-pan;
 \\
Joint Institute for Nuclear Research, Dubna;
 \\
National Research Foundation of Korea (NRF);
 \\
CONACYT, DGAPA, M\'{e}xico, ALFA-EC and the EPLANET Program
(European Particle Physics Latin American Network)
 \\
Stichting voor Fundamenteel Onderzoek der Materie (FOM) and the Nederlandse Organisatie voor Wetenschappelijk Onderzoek (NWO), Netherlands;
 \\
Research Council of Norway (NFR);
 \\
Polish Ministry of Science and Higher Education;
 \\
National Science Centre, Poland;
 \\
 Ministry of National Education/Institute for Atomic Physics and CNCS-UEFISCDI - Romania;
 \\
Ministry of Education and Science of Russian Federation, Russian
Academy of Sciences, Russian Federal Agency of Atomic Energy,
Russian Federal Agency for Science and Innovations and The Russian
Foundation for Basic Research;
 \\
Ministry of Education of Slovakia;
 \\
Department of Science and Technology, South Africa;
 \\
CIEMAT, EELA, Ministerio de Econom\'{i}a y Competitividad (MINECO) of Spain, Xunta de Galicia (Conseller\'{\i}a de Educaci\'{o}n),
CEA\-DEN, Cubaenerg\'{\i}a, Cuba, and IAEA (International Atomic Energy Agency);
 \\
Swedish Research Council (VR) and Knut $\&$ Alice Wallenberg
Foundation (KAW);
 \\
Ukraine Ministry of Education and Science;
 \\
United Kingdom Science and Technology Facilities Council (STFC);
 \\
The United States Department of Energy, the United States National
Science Foundation, the State of Texas, and the State of Ohio.

%% file: Alice_Authorlist_2013-Dec-03-CERNPREP.tex


\begingroup
\small
\begin{flushleft}
B.~Abelev\Irefn{org72}\And
J.~Adam\Irefn{org37}\And
D.~Adamov\'{a}\Irefn{org80}\And
M.M.~Aggarwal\Irefn{org84}\And
G.~Aglieri~Rinella\Irefn{org34}\And
M.~Agnello\Irefn{org91}\textsuperscript{,}\Irefn{org108}\And
A.~Agostinelli\Irefn{org26}\And
N.~Agrawal\Irefn{org44}\And
Z.~Ahammed\Irefn{org127}\And
N.~Ahmad\Irefn{org18}
% ...
\renewcommand\labelenumi{\textsuperscript{\theenumi}~}

\section*{Affiliation notes}
\renewcommand\theenumi{\roman{enumi}}
\begin{Authlist}
\item \Adef{0}Deceased
\item \Adef{idp1129376}{Also at: St. Petersburg State Polytechnical University}
\item \Adef{idp3041504}{Also at: Department of Applied Physics, Aligarh Muslim University, Aligarh, India}
\item \Adef{idp3747728}{Also at: M.V. Lomonosov Moscow State University, D.V. Skobeltsyn Institute of Nuclear Physics, Moscow, Russia}
\item \Adef{idp4004592}{Also at: University of Belgrade, Faculty of Physics and "Vin\v{c}a" Institute of Nuclear Sciences, Belgrade, Serbia}
\item \Adef{idp4287040}{Permanent Address: Permanent Address: Konkuk University, Seoul, Korea}
\item \Adef{idp4812160}{Also at: Institute of Theoretical Physics, University of Wroclaw, Wroclaw, Poland}
\item \Adef{idp5727408}{Also at: University of Kansas, Lawrence, KS, United States}
\end{Authlist}

\section*{Collaboration Institutes}
\renewcommand\theenumi{\arabic{enumi}~}
\begin{Authlist}

\item \Idef{org1}A.I. Alikhanyan National Science Laboratory (Yerevan Physics Institute) Foundation, Yerevan, Armenia
\item \Idef{org2}Benem\'{e}rita Universidad Aut\'{o}noma de Puebla, Puebla, Mexico
\item \Idef{org3}Bogolyubov Institute for Theoretical Physics, Kiev, Ukraine
\item \Idef{org4}Bose Institute, Department of Physics and Centre for Astroparticle Physics and Space Science (CAPSS), Kolkata, India
\item \Idef{org5}Budker Institute for Nuclear Physics, Novosibirsk, Russia
\item \Idef{org6}California Polytechnic State University, San Luis Obispo, CA, United States
\item \Idef{org7}Central China Normal University, Wuhan, China
\item \Idef{org8}Centre de Calcul de l'IN2P3, Villeurbanne, France
\item \Idef{org9}Centro de Aplicaciones Tecnol\'{o}gicas y Desarrollo Nuclear (CEADEN), Havana, Cuba
\item \Idef{org10}Centro de Investigaciones Energ\'{e}ticas Medioambientales y Tecnol\'{o}gicas (CIEMAT), Madrid, Spain
\item \Idef{org11}Centro de Investigaci\'{o}n y de Estudios Avanzados (CINVESTAV), Mexico City and M\'{e}rida, Mexico
\item \Idef{org12}Centro Fermi - Museo Storico della Fisica e Centro Studi e Ricerche ``Enrico Fermi'', Rome, Italy
\item \Idef{org13}Chicago State University, Chicago, USA
\item \Idef{org14}Commissariat \`{a} l'Energie Atomique, IRFU, Saclay, France
\item \Idef{org15}COMSATS Institute of Information Technology (CIIT), Islamabad, Pakistan
\item \Idef{org16}Departamento de F\'{\i}sica de Part\'{\i}culas and IGFAE, Universidad de Santiago de Compostela, Santiago de Compostela, Spain
\item \Idef{org17}Department of Physics and Technology, University of Bergen, Bergen, Norway
\item \Idef{org18}Department of Physics, Aligarh Muslim University, Aligarh, India
\item \Idef{org19}Department of Physics, Ohio State University, Columbus, OH, United States
\item \Idef{org20}Department of Physics, Sejong University, Seoul, South Korea
\item \Idef{org21}Department of Physics, University of Oslo, Oslo, Norway
\item \Idef{org22}Dipartimento di Fisica dell'Universit\`{a} 'La Sapienza' and Sezione INFN Rome, Italy
\item \Idef{org23}Dipartimento di Fisica dell'Universit\`{a} and Sezione INFN, Cagliari, Italy
\item \Idef{org24}Dipartimento di Fisica dell'Universit\`{a} and Sezione INFN, Trieste, Italy
\item \Idef{org25}Dipartimento di Fisica dell'Universit\`{a} and Sezione INFN, Turin, Italy
\item \Idef{org26}Dipartimento di Fisica e Astronomia dell'Universit\`{a} and Sezione INFN, Bologna, Italy
\item \Idef{org27}Dipartimento di Fisica e Astronomia dell'Universit\`{a} and Sezione INFN, Catania, Italy
\item \Idef{org28}Dipartimento di Fisica e Astronomia dell'Universit\`{a} and Sezione INFN, Padova, Italy
\item \Idef{org29}Dipartimento di Fisica `E.R.~Caianiello' dell'Universit\`{a} and Gruppo Collegato INFN, Salerno, Italy
\item \Idef{org30}Dipartimento di Scienze e Innovazione Tecnologica dell'Universit\`{a} del  Piemonte Orientale and Gruppo Collegato INFN, Alessandria, Italy
\item \Idef{org31}Dipartimento Interateneo di Fisica `M.~Merlin' and Sezione INFN, Bari, Italy
\item \Idef{org32}Division of Experimental High Energy Physics, University of Lund, Lund, Sweden
\item \Idef{org33}Eberhard Karls Universit\"{a}t T\"{u}bingen, T\"{u}bingen, Germany
\item \Idef{org34}European Organization for Nuclear Research (CERN), Geneva, Switzerland
\item \Idef{org35}Faculty of Engineering, Bergen University College, Bergen, Norway
\item \Idef{org36}Faculty of Mathematics, Physics and Informatics, Comenius University, Bratislava, Slovakia
\item \Idef{org37}Faculty of Nuclear Sciences and Physical Engineering, Czech Technical University in Prague, Prague, Czech Republic
\item \Idef{org38}Faculty of Science, P.J.~\v{S}af\'{a}rik University, Ko\v{s}ice, Slovakia
\item \Idef{org39}Frankfurt Institute for Advanced Studies, Johann Wolfgang Goethe-Universit\"{a}t Frankfurt, Frankfurt, Germany
\item \Idef{org40}Gangneung-Wonju National University, Gangneung, South Korea
\item \Idef{org41}Gauhati University, Department of Physics, Guwahati, India
\item \Idef{org42}Helsinki Institute of Physics (HIP), Helsinki, Finland
\item \Idef{org43}Hiroshima University, Hiroshima, Japan
\item \Idef{org44}Indian Institute of Technology Bombay (IIT), Mumbai, India
\item \Idef{org45}Indian Institute of Technology Indore, Indore (IITI), India
\item \Idef{org46}Inha University, Incheon, South Korea
\item \Idef{org47}Institut de Physique Nucl\'eaire d'Orsay (IPNO), Universit\'e Paris-Sud, CNRS-IN2P3, Orsay, France
\item \Idef{org48}Institut f\"{u}r Informatik, Johann Wolfgang Goethe-Universit\"{a}t Frankfurt, Frankfurt, Germany
\item \Idef{org49}Institut f\"{u}r Kernphysik, Johann Wolfgang Goethe-Universit\"{a}t Frankfurt, Frankfurt, Germany
\item \Idef{org50}Institut f\"{u}r Kernphysik, Technische Universit\"{a}t Darmstadt, Darmstadt, Germany
\item \Idef{org51}Institut f\"{u}r Kernphysik, Westf\"{a}lische Wilhelms-Universit\"{a}t M\"{u}nster, M\"{u}nster, Germany
\item \Idef{org52}Institut Pluridisciplinaire Hubert Curien (IPHC), Universit\'{e} de Strasbourg, CNRS-IN2P3, Strasbourg, France
\item \Idef{org53}Institute for Nuclear Research, Academy of Sciences, Moscow, Russia
\item \Idef{org54}Institute for Subatomic Physics of Utrecht University, Utrecht, Netherlands
\item \Idef{org55}Institute for Theoretical and Experimental Physics, Moscow, Russia
\item \Idef{org56}Institute of Experimental Physics, Slovak Academy of Sciences, Ko\v{s}ice, Slovakia
\item \Idef{org57}Institute of Physics, Academy of Sciences of the Czech Republic, Prague, Czech Republic
\item \Idef{org58}Institute of Physics, Bhubaneswar, India
\item \Idef{org59}Institute of Space Science (ISS), Bucharest, Romania
\item \Idef{org60}Instituto de Ciencias Nucleares, Universidad Nacional Aut\'{o}noma de M\'{e}xico, Mexico City, Mexico
\item \Idef{org61}Instituto de F\'{\i}sica, Universidad Nacional Aut\'{o}noma de M\'{e}xico, Mexico City, Mexico
\item \Idef{org62}iThemba LABS, National Research Foundation, Somerset West, South Africa
\item \Idef{org63}Joint Institute for Nuclear Research (JINR), Dubna, Russia
\item \Idef{org64}Konkuk University, Seoul, South Korea
\item \Idef{org65}Korea Institute of Science and Technology Information, Daejeon, South Korea
\item \Idef{org66}KTO Karatay University, Konya, Turkey
\item \Idef{org67}Laboratoire de Physique Corpusculaire (LPC), Clermont Universit\'{e}, Universit\'{e} Blaise Pascal, CNRS--IN2P3, Clermont-Ferrand, France
\item \Idef{org68}Laboratoire de Physique Subatomique et de Cosmologie (LPSC), Universit\'{e} Joseph Fourier, CNRS-IN2P3, Institut Polytechnique de Grenoble, Grenoble, France
\item \Idef{org69}Laboratori Nazionali di Frascati, INFN, Frascati, Italy
\item \Idef{org70}Laboratori Nazionali di Legnaro, INFN, Legnaro, Italy
\item \Idef{org71}Lawrence Berkeley National Laboratory, Berkeley, CA, United States
\item \Idef{org72}Lawrence Livermore National Laboratory, Livermore, CA, United States
\item \Idef{org73}Moscow Engineering Physics Institute, Moscow, Russia
\item \Idef{org74}National Centre for Nuclear Studies, Warsaw, Poland
\item \Idef{org75}National Institute for Physics and Nuclear Engineering, Bucharest, Romania
\item \Idef{org76}National Institute of Science Education and Research, Bhubaneswar, India
\item \Idef{org77}Niels Bohr Institute, University of Copenhagen, Copenhagen, Denmark
\item \Idef{org78}Nikhef, National Institute for Subatomic Physics, Amsterdam, Netherlands
\item \Idef{org79}Nuclear Physics Group, STFC Daresbury Laboratory, Daresbury, United Kingdom
\item \Idef{org80}Nuclear Physics Institute, Academy of Sciences of the Czech Republic, \v{R}e\v{z} u Prahy, Czech Republic
\item \Idef{org81}Oak Ridge National Laboratory, Oak Ridge, TN, United States
\item \Idef{org82}Petersburg Nuclear Physics Institute, Gatchina, Russia
\item \Idef{org83}Physics Department, Creighton University, Omaha, NE, United States
\item \Idef{org84}Physics Department, Panjab University, Chandigarh, India
\item \Idef{org85}Physics Department, University of Athens, Athens, Greece
\item \Idef{org86}Physics Department, University of Cape Town, Cape Town, South Africa
\item \Idef{org87}Physics Department, University of Jammu, Jammu, India
\item \Idef{org88}Physics Department, University of Rajasthan, Jaipur, India
\item \Idef{org89}Physik Department, Technische Universit\"{a}t M\"{u}nchen, Munich, Germany
\item \Idef{org90}Physikalisches Institut, Ruprecht-Karls-Universit\"{a}t Heidelberg, Heidelberg, Germany
\item \Idef{org91}Politecnico di Torino, Turin, Italy
\item \Idef{org92}Purdue University, West Lafayette, IN, United States
\item \Idef{org93}Pusan National University, Pusan, South Korea
\item \Idef{org94}Research Division and ExtreMe Matter Institute EMMI, GSI Helmholtzzentrum f\"ur Schwerionenforschung, Darmstadt, Germany
\item \Idef{org95}Rudjer Bo\v{s}kovi\'{c} Institute, Zagreb, Croatia
\item \Idef{org96}Russian Federal Nuclear Center (VNIIEF), Sarov, Russia
\item \Idef{org97}Russian Research Centre Kurchatov Institute, Moscow, Russia
\item \Idef{org98}Saha Institute of Nuclear Physics, Kolkata, India
\item \Idef{org99}School of Physics and Astronomy, University of Birmingham, Birmingham, United Kingdom
\item \Idef{org100}Secci\'{o}n F\'{\i}sica, Departamento de Ciencias, Pontificia Universidad Cat\'{o}lica del Per\'{u}, Lima, Peru
\item \Idef{org101}Sezione INFN, Bari, Italy
\item \Idef{org102}Sezione INFN, Bologna, Italy
\item \Idef{org103}Sezione INFN, Cagliari, Italy
\item \Idef{org104}Sezione INFN, Catania, Italy
\item \Idef{org105}Sezione INFN, Padova, Italy
\item \Idef{org106}Sezione INFN, Rome, Italy
\item \Idef{org107}Sezione INFN, Trieste, Italy
\item \Idef{org108}Sezione INFN, Turin, Italy
\item \Idef{org109}SSC IHEP of NRC "Kurchatov institute" , Protvino, Russia
\item \Idef{org110}SUBATECH, Ecole des Mines de Nantes, Universit\'{e} de Nantes, CNRS-IN2P3, Nantes, France
\item \Idef{org111}Suranaree University of Technology, Nakhon Ratchasima, Thailand
\item \Idef{org112}Technical University of Split FESB, Split, Croatia
\item \Idef{org113}The Henryk Niewodniczanski Institute of Nuclear Physics, Polish Academy of Sciences, Cracow, Poland
\item \Idef{org114}The University of Texas at Austin, Physics Department, Austin, TX, USA
\item \Idef{org115}Universidad Aut\'{o}noma de Sinaloa, Culiac\'{a}n, Mexico
\item \Idef{org116}Universidade de S\~{a}o Paulo (USP), S\~{a}o Paulo, Brazil
\item \Idef{org117}Universidade Estadual de Campinas (UNICAMP), Campinas, Brazil
\item \Idef{org118}University of Houston, Houston, TX, United States
\item \Idef{org119}University of Jyv\"{a}skyl\"{a}, Jyv\"{a}skyl\"{a}, Finland
\item \Idef{org120}University of Liverpool, Liverpool, United Kingdom
\item \Idef{org121}University of Tennessee, Knoxville, TN, United States
\item \Idef{org122}University of Tokyo, Tokyo, Japan
\item \Idef{org123}University of Tsukuba, Tsukuba, Japan
\item \Idef{org124}University of Zagreb, Zagreb, Croatia
\item \Idef{org125}Universit\'{e} de Lyon, Universit\'{e} Lyon 1, CNRS/IN2P3, IPN-Lyon, Villeurbanne, France
\item \Idef{org126}V.~Fock Institute for Physics, St. Petersburg State University, St. Petersburg, Russia
\item \Idef{org127}Variable Energy Cyclotron Centre, Kolkata, India
\item \Idef{org128}Vestfold University College, Tonsberg, Norway
\item \Idef{org129}Warsaw University of Technology, Warsaw, Poland
\item \Idef{org130}Wayne State University, Detroit, MI, United States
\item \Idef{org131}Wigner Research Centre for Physics, Hungarian Academy of Sciences, Budapest, Hungary
\item \Idef{org132}Yale University, New Haven, CT, United States
\item \Idef{org133}Yonsei University, Seoul, South Korea
\item \Idef{org134}Zentrum f\"{u}r Technologietransfer und Telekommunikation (ZTT), Fachhochschule Worms, Worms, Germany
\end{Authlist}
\endgroup